\documentclass{PoS}

\usepackage{subcaption} % side by side figures etc
\usepackage{multirow} % multirow/column tables
\usepackage{siunitx} % units, angles etc. - does not work with deluxetable ... :(
\newcommand*\Fermis{\textit{Fermi}-LAT }

\newcommand{\similar}{\ensuremath{\sim}}

\newcommand{\GR}{$\gamma$-ray}
\newcommand{\GRs}{$\gamma$-ray }
\newcommand*\degrees{$^{\circ}$ }

\title{VERITAS Observations Of M~31 (The Andromeda Galaxy)}

\ShortTitle{VERITAS Observations Of M~31 (The Andromeda Galaxy)}

\author{\speaker{Ralph Bird}, for the VERITAS Collaboration\thanks{http://veritas.sao.arizona.edu}\\
        University College Dublin\\
        E-mail: \email{ralph.bird.1@gmail.com}}

\abstract{Diffuse gamma rays are tracers of cosmic rays, providing information on their origin, interaction and diffusion through a galaxy. M 31 (the Andromeda Galaxy) is the closest spiral galaxy to the Milky Way (d = \SI{780}{kpc}) and is very well studied at all wavelengths. Thus it is a prime target for the study of diffuse gamma-ray emission. The very-high-energy (VHE, E~$>$~\SI{100}{GeV}) gamma-ray observatory VERITAS has conducted 54 hours of observations of M~31 and an upper limit on the VHE flux is presented along with an updated \Fermis ({0.1}~$<$~E~$<$~\SI{300}{GeV}) analysis. These observations will be compared with predictions of the gamma-ray flux derived from models of the inelastic scattering of VHE cosmic rays of the interstellar medium (ISM) and the interstellar radiation field. M 31 provides an ideal opportunity to probe this mechanism. Its proximity and spatial extent, significantly larger than the VERITAS point spread function but smaller than the field-of-view, potentially enables the star-forming ring, \SI{10}{kpc} from the galaxy core, with its dense ISM and numerous supernova remnants to be resolved.}

\FullConference{The 34th International Cosmic Ray Conference,\\
		30 July- 6 August, 2015\\
		The Hague, The Netherlands}

\begin{document}

\section{Gamma-Ray Emission in M~31}
M~31 is the closest spiral galaxy to the Milky Way (MW) with apparent dimensions of 3.2\degrees by 1\degrees \cite{NED} allowing its internal structure to be studied in detail. 
Of particular interest is a gas-rich, star-forming ring \similar\SI{10}{kpc} from the galaxy centre.
M~31 has been detected by the \Fermis with an integral photon flux ($>$\SI{100}{MeV}) of $(9.1 \pm 1.9_{stat} \pm 1.0_{sys}) \times 10^{-9} \si{ph cm^{-2} s^{-1}}$ as a point source but with some evidence of spatial extension  \cite{Abdo2010}.
M~31 has not been detected in very-high-energy (VHE) $\gamma$-rays, upper limits on the VHE flux were presented by the HEGRA collaboration in \cite{Aharonian2003} at 3.3\% of the Crab Nebula flux for individual point sources close to the center of the galaxy rising to $\sim$30\% near the edges.

There are expected to be two main contributors to the VHE emission from M~31; diffuse emission from the interaction of cosmic rays with the interstellar medium (ISM) and unresolved point-source emission.
Due to its apparent size, a detection by VERITAS will allow it to map the VHE \GRs emission and thus we could determine where the emission is coming from.
It could be regions where there are a number of potentially unresolved point sources, suggesting that it is the sources themselves, or cosmic rays that are only diffusing a short distance from the source population, that are producing the signal.
If the emission is more diffuse and coming from regions away from potential sources, then it will suggest that the emission is caused by cosmic-rays diffusing large distances through the galaxy from their sources.
Since hadronic cosmic rays have a significantly longer lifetime than leptonic cosmic rays this would provide good evidence for a hadronic origin.
The \GRs emission spectrum will also provide information about the underlying physics, in particular at around \SI{0.1}{GeV} where detection of a low energy cut-off (a ``pion-bump'') would provide significant evidence that the diffuse \GRs emission is of hadronic origin.

\section{Predicted Gamma-Ray Emission}
\begin{figure}[h]
\centering
\includegraphics[width=0.8\linewidth]{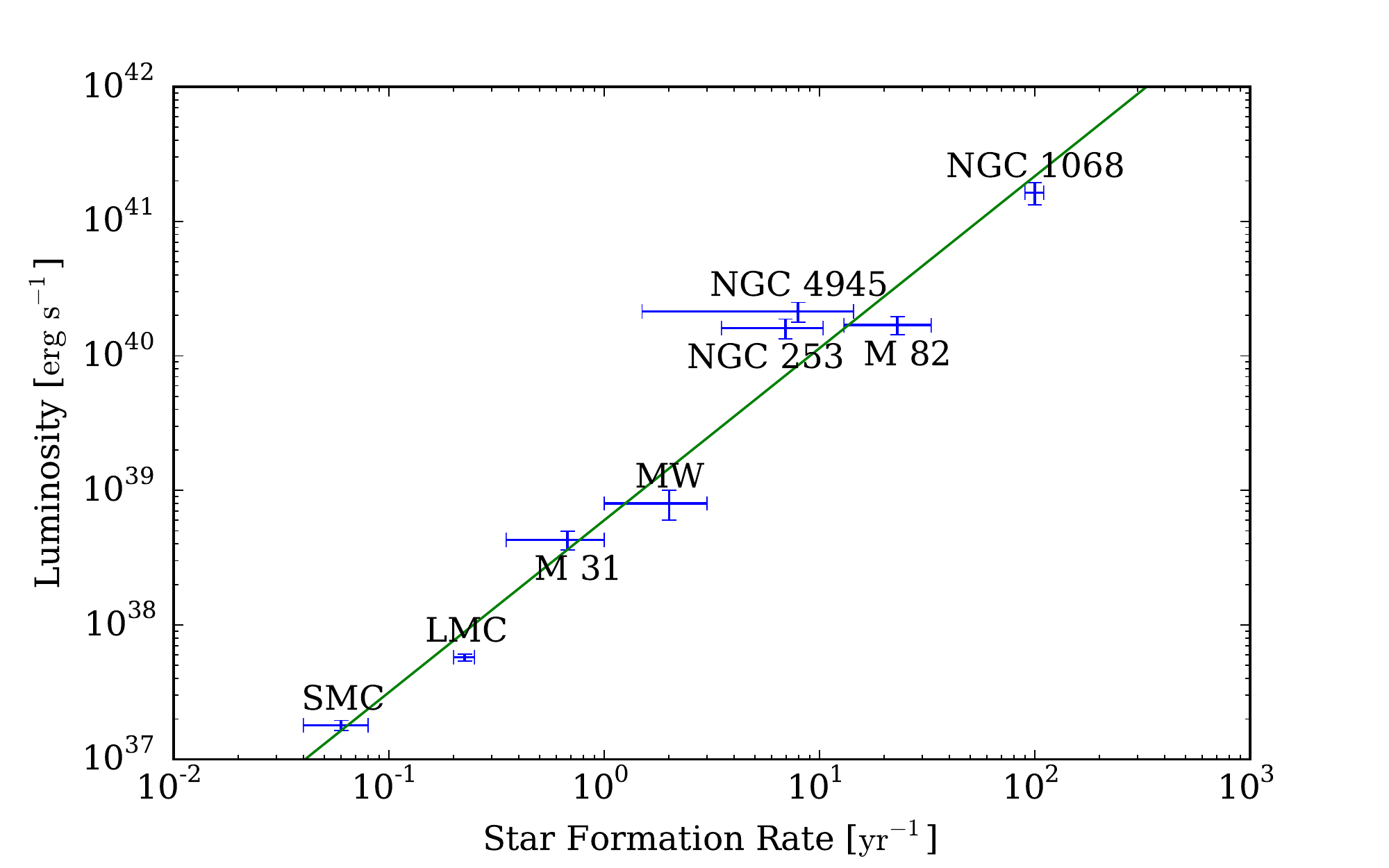}
\caption[\GRs Luminosity vs. SFR]{Gamma ray luminosity (0.1 to \SI{100}{GeV}) vs. SFR for ``normal'' galaxies overlaid with the best fit, $\mathrm{Luminosity} \propto (\mathrm{SFR})^{1.28}$}
\label{Fig:LumSFR}
\end{figure}

It is generally understood that the \GRs luminosity of  normal galaxies (that is, galaxies without an active galactic nucleus) scales with a few key parameters; the number of cosmic ray accelerators (which scales with the star formation rate (SFR) \cite{Condon1992}), the escape time of the cosmic rays  and the amount of target material.
Working from these assumptions a number of predictions have been made on the HE \GRs flux, for example \cite{Pavlidou2001a}  predicts the flux using the supernova rate and the mass of hydrogen.
This relationship has been explored using existing flux measurements taken from the 3FGL catalogue \cite{Fermi2015} for normal and starburst galaxies, and the two Seyfert galaxies NGC~4945 and NGC~1068  (it is assumed that the emission is dominated by the star forming regions rather than the central black hole).
There is quite a strong correlation, as shown in Figure~\ref{Fig:LumSFR}, with the best fit showing that the luminosity scales with $SFR^{1.28}$ rather than the $SFR^{1}$ used in \cite{Pavlidou2001a}. 
However, in \cite{Pavlidou2001a} the luminosity also depends upon the mass of hydrogen and since the SFR also depends upon the mass of hydrogen a relationship of $SFR^{1.28}$ is not unexpected \cite{Persic2011}. 
Using this scaling relationship and the measurements of the VHE flux from the starburst galaxies M~82 \cite{Acciari2009} and NGC~253 \cite{Abramowski2012} allows for predictions of the M~31 integral flux to be made of \SI{8.79e-14}{cm^{-2} s^{-1}} (above \SI{750}{GeV}) and \SI{7.24e-13}{cm^{-2} s^{-1}} (above \SI{190}{GeV}) respectively. 

M~31 is often considered a sister galaxy to the Milky Way and thus, to a reasonable approximation, a model derived for one galaxy can be used to model the other.
GALPROP \cite{GALPROP} provides a very detailed model of the expected \GRs emission from the Milky Way (no equivalent models are available for M~31). 
Using the scaling relationship and the relative SFRs ($1 < \mathrm{SFR}_{MW}<3, \; 0.35< \mathrm{SFR}_{M31} < 1, \; \frac{\mathrm{SFR}_{M31}}{\mathrm{SFR}_{MW}} \approx 0.5$ \cite{Abdo2010}), and placing the Milky Way at the distance of M~31 (\SI{780}{kpc}), an estimate of the M~31 \GRs flux can be determined.  
In this work the model z04LMS from \cite{Strong2010} is used and scaled by a factor of $0.5^{1.28} = 0.41$, hereafter this is referred to as the GALPROP model.

\section{Updated \textit{Fermi}-LAT Analysis}
\begin{sloppypar}Following its the initial detection \cite{Abdo2010}, M~31 has appeared in both the 2FGL catalogue (J0042.5+4114 \cite{Nolan2012}) and the 3FGL catalogue (J0042.5+4117 \cite{Fermi2015}), in both cases as a point source.
Analyses have also been published in \cite{Li2013} where they examine dark matter limits from M~31 and in \cite{Pshirkov2015} where  evidence for \GRs emission from a spatially extended halo around M~31 is presented.
We have conducted an updated analysis of the \Fermis data, using the \textit{LATAnalysisScripts} and \textit{Fermi Science Tools v9r33p0} to conduct a binned (\ang{0.1}) analysis.
Six and a half years of Pass 7 data covering the energy range $0.1-\SI{300}{GeV}$ were analysed within \ang{30} of the center of M~31 ((RA, Dec (J2000)) = (0$^h$ 43$^{m}$ 35$^s$.43, +\ang{41} 20' 56".8)) with a \ang{15} ROI.
The \textit{P7REP\_SOURCE\_V15} IRFs were used for event class 2 with quality cuts of  DATA\_QUAL==1, LAT\_CONFIG==1 and ABS(ROCK\_ANGLE)$<$52.
\end{sloppypar}

\begin{figure}[h]
\centering
\hspace{1cm}
\begin{subfigure}[t]{0.45\textwidth}
\includegraphics[width=1\linewidth]{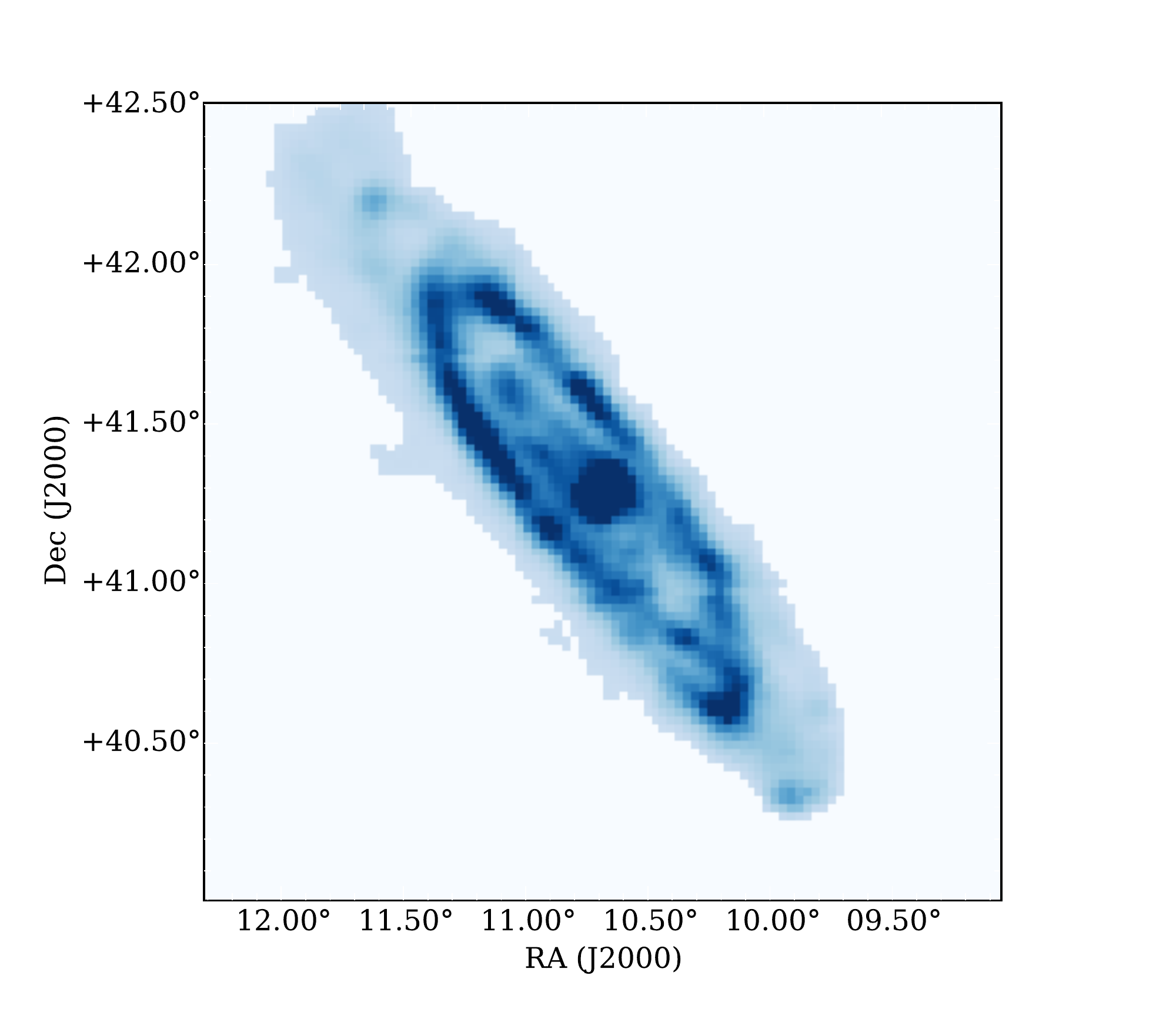}
\caption{\textit{Raw} Template}
\label{Fig:IRISRaw}
\end{subfigure}
\hfill
\begin{subfigure}[t]{0.45\textwidth}
\includegraphics[width=1\linewidth]{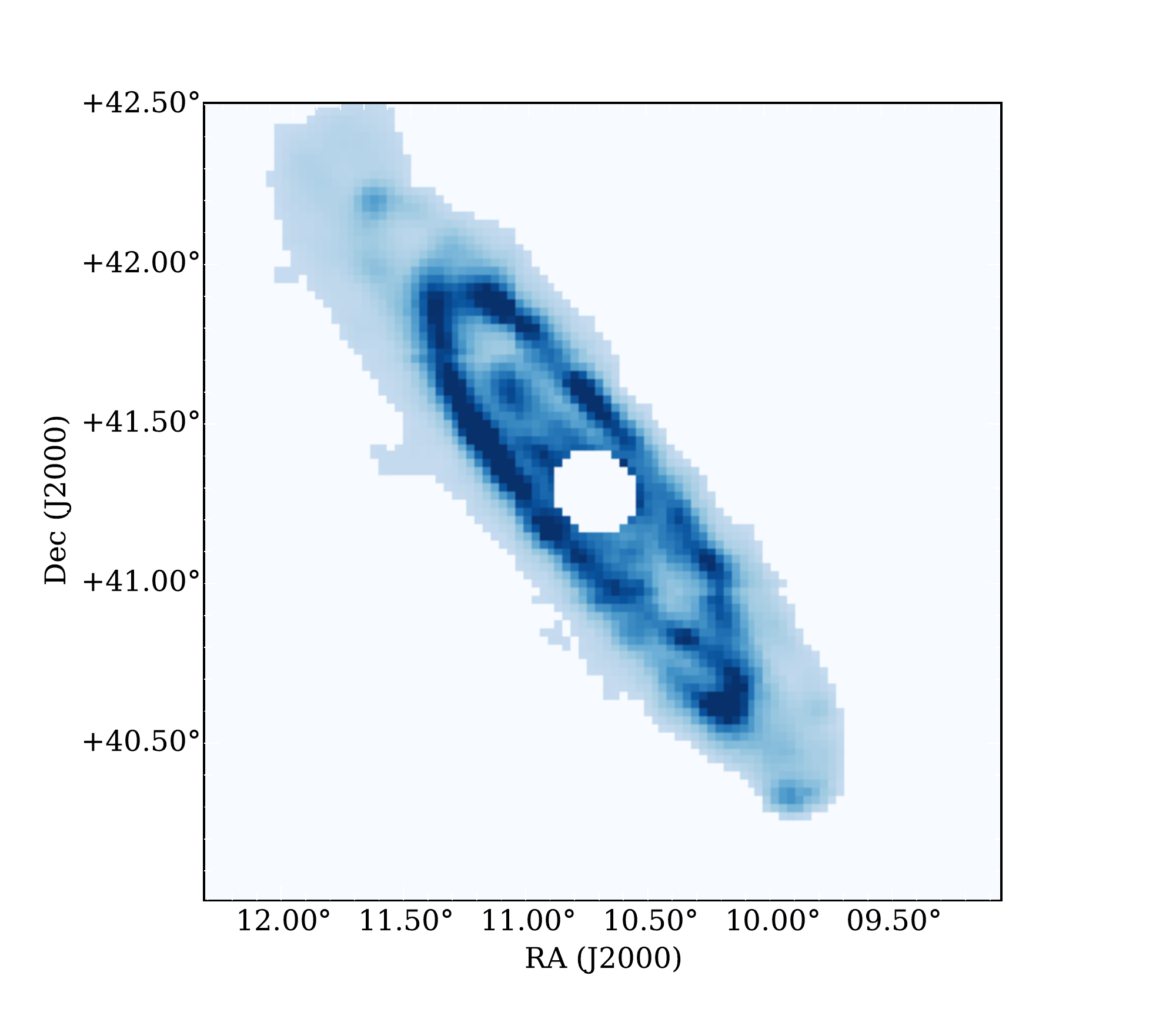}
\caption{\textit{Ring} Template}
\label{Fig:IRISRing}
\end{subfigure}
\hspace{1cm}
\caption{The extended templates used in the \Fermis analysis derived from the IRAS Survey (IRIS) 100~$\mu$m far infrared map of M~31 \cite{MivilleDeschenes2005}}
\end{figure}

Four different models of M~31 were tested: \textbf{Point} - a point source as published in the 3FGL. 
\textbf{Raw} - a template to test for extended emission generated using the same Improved Reprocessing of the IRAS Survey (IRIS) \SI{100}{\micro m} far infrared map (Figure~\ref{Fig:IRISRaw}) that was used in the original \Fermis detection paper \cite{Abdo2010}. 
\textbf{Ring} - the \textit{Raw} template but with the central ``blob'' removed (Figure~\ref{Fig:IRISRing}) to test whether emission is better fit with a ring structure as seen in the hydrogen maps. \textbf{Both} - the \textit{Ring} and the \textit{Point} templates in the same fit, this is to test the relative importances of the two models.
All four models were fit as a single powerlaw with the index and normalisation (at \SI{0.57278}{GeV}) free to vary.

The background model was generated using the galactic diffuse model \textit{gll\_iem\_v05\_rev1}, the isotropic diffuse model \textit{iso\_source\_v05} and point sources from the 3FGL catalogue (all sources outside the ROI (\ang{15}) were fixed to the values from the 3FGL).
Following the production of a TS map of the central \ang{7.5} region, an additional point source at  ((RA, Dec (J2000)) = (0$^h$ 40$^m$ 53$^s$.43,  +\ang{42} 15' 02".0)) was added to the background model as a power law point source with the index and amplitude left free to vary. 
All four models were initially fit with a power law spectrum with the normalisation energy from the 3FGL, \SI{0.57278}{GeV} (Table~\ref{Tab:FermiTS}).
These results show that there is marginal evidence that M~31 is best fit by an extended source, either the \textit{Raw} or the \textit{Ring} model (at the 2.24 and 1.92$\sigma$ level respectively), whether this is a reflection of the nature of the source or the PSF at low energies ($>$\ang{1} for energies less than \SI{1}{GeV}) which is comparable to the galaxy size needs further investigation.

\begin{table}[h]
\centering
\begin{tabular}{ccccc}
\hline
\multicolumn{2}{c}{\multirow{2}{*}{Model}} & \multirow{2}{*}{TS} & Prefactor /  & \multirow{2}{*}{Index} \\
& & & $\times 10^{-12}$cm$^{-2}$ s$^{-1}$ MeV$^{-1}$ & \\
\hline
\multicolumn{2}{c}{Point} & 59.28 & 1.90 $\pm$ 0.06 & -2.50 $\pm$ 0.02\\
\multicolumn{2}{c}{Raw}   & 64.34 & 2.51 $\pm$ 0.41 & -2.23 $\pm$ 0.10 \\
\multicolumn{2}{c}{Ring}  & 62.98 & 2.51 $\pm$ 0.43 & -2.21 $\pm$ 0.10 \\
\multirow{2}{*}{Both} & (Ring)  & 19.94 & 1.36 $\pm$ 0.65 & -2.14 $\pm$ 0.17 \\
 & (Point) & 17.94 & 1.01 $\pm$ 0.18 & -2.42 $\pm$ 0.18 \\
\hline
\end{tabular}
\caption[M~31 \Fermis Analysis Results]{Results of the \Fermis analysis of the M~31 region.}
\label{Tab:FermiTS}
\end{table}

\begin{figure}[h]
\centering
\includegraphics[width=0.7\linewidth]{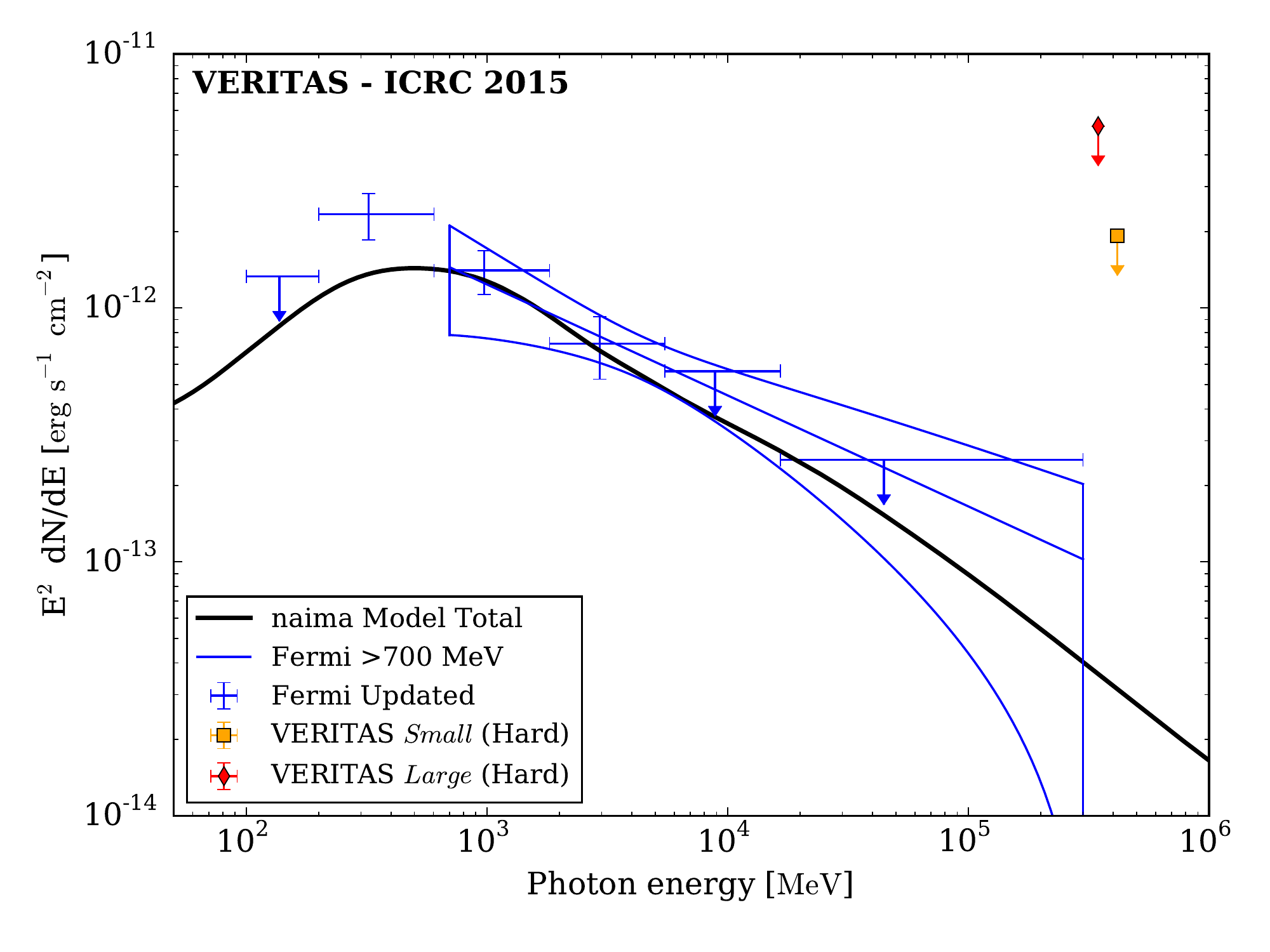}
\caption{The VERITAS upper limits compared with the \Fermis spectra and the GALPROP model.  All upper limits are at 95\%.}
\label{Fig:VERM31ULcomp}
\end{figure}

For the \textit{Raw} model a spectrum for M~31 was calculated using the same binning as that used in the original detection paper with an additional bin covering $0.1 - \SI{0.2}{GeV}$ and with the highest energy bin extended from $16.572-\SI{50}{GeV}$ to $16.572-\SI{300}{GeV}$.
Looking at the GALPROP model (Figure~\ref{Fig:VERM31ULcomp}) we expect curvature at low energies and thus a power law fit over the whole range is not suitable. 
Instead,  a complete reanalysis was conducted with a low energy threshold of \SI{0.7}{GeV}, this gives a spectrum with a differential flux of $(2.11 \pm 0.40_{\mathrm{stat}}) \times 10^{-14} \si{cm^{-2} s^{-1} MeV^{-1}}$ at a decorrelation energy of \SI{4.384}{GeV} and with a power law index of $-(2.44 \pm 0.23_{\mathrm{stat}})$.
The $0.7-\SI{300}{GeV}$ flux is $(9.0 \pm 1.7_{\mathrm{stat}})\times10^{-10} \si{photons cm^{-2} s^{-1}}$.

Overlaying these data points with the GALPROP model shows a good agreement with the points (Figure~\ref{Fig:VERM31ULcomp}).
The upper limit in the $0.1-\SI{0.2}{GeV}$ range is highly suggestive of a ``pion bump'', though, with the current level of detail it is hard to draw a firm conclusion. 
The release of \textit{Pass 8} with its significant improvements at low energies will provide further insight into this.

\section{VERITAS Observations and Results}
VERITAS is an array of four imaging atmospheric Cherenkov telescopes (IACT) located at the Fred Lawrence Whipple Observatory (FLWO) in southern Arizona (\ang{31} 40'N, \ang{110} 57'W, \SI{1.3}{km} a.s.l.).
Designed to detect the Cherenkov emission from extensive air showers produced by cosmic and \GR s, each telescope has a mirror area of \SI{110}{m^2} and is equipped with a 499-pixel camera of \ang{3.5} diameter field-of-view (FoV) with an angular resolution of \ang{0.1} at \SI{1}{TeV}. 
The system, completed in 2007, is run in a coincident mode requiring at least two of the four telescopes to trigger in each event. 
This design enables the observations of astrophysical objects in the energy range from \SI{85}{GeV} to $>$\SI{30}{TeV}. 
For VERITAS, M~31 is a spatially extended, optically bright target, presenting significant challenges for analysis since the standard  techniques are optimised for point sources in optically dark sky regions \cite{Daniel2007}.
To overcome the optical brightness (which impacts upon the detection of the Cherenkov emission from the extensive air shower) a large \emph{Size} (number of digital counts in the cleaned image) cut was employed prior to the reconstruction of the images at the cost of an increase in the energy threshold.
With this cut in place the impact of the optical brightness (tested using events that fail the gamma/hadron selection cuts) is restricted to a small central region which is excluded from this analysis.

\begin{figure}[h]
\centering
\begin{subfigure}[t]{0.49\textwidth}
\includegraphics[width=1\linewidth]{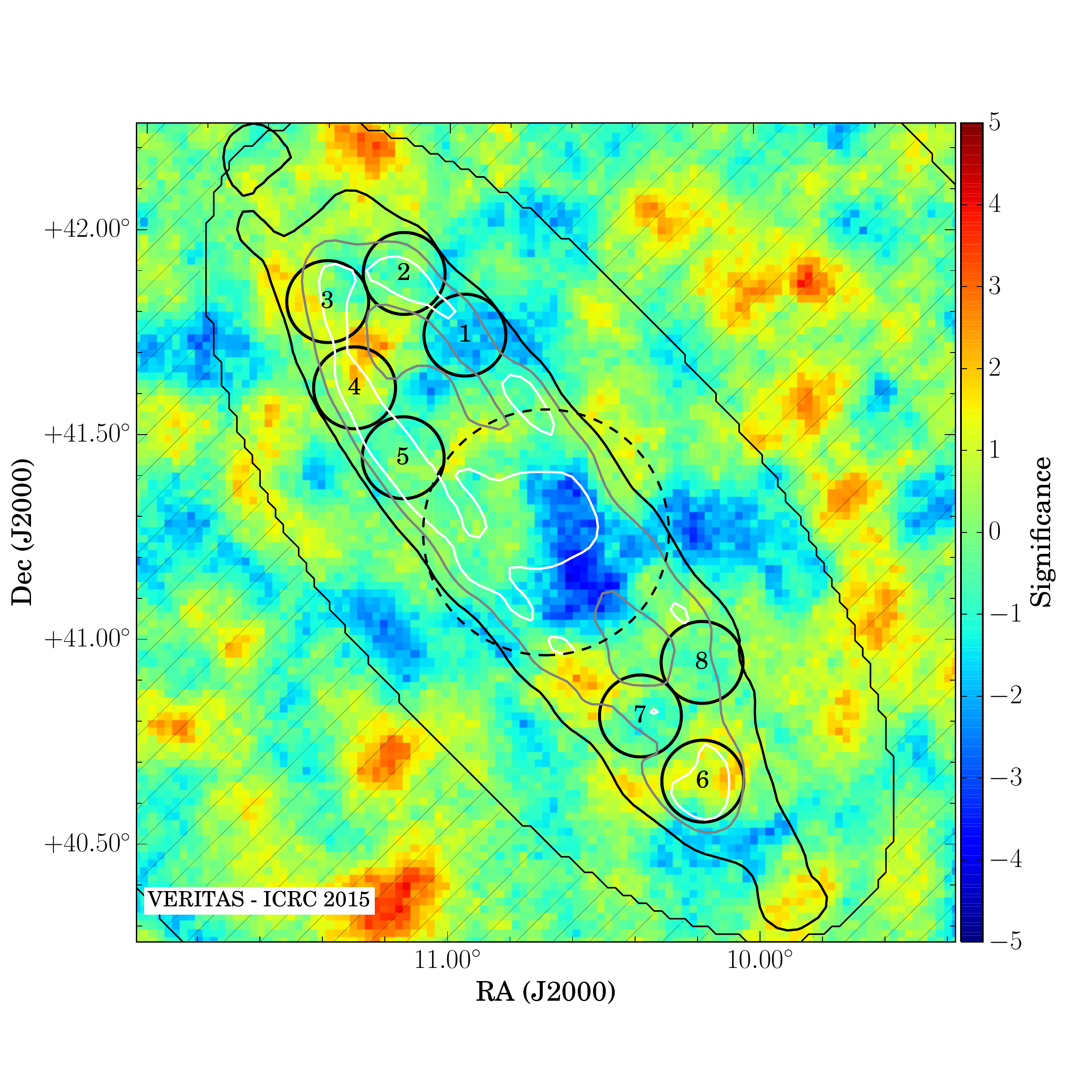}
\caption{test region radius = \ang{0.1} (\textit{Small})}
\label{Fig:VERSkyMapAllHardED}
\end{subfigure}
\hfill
\begin{subfigure}[t]{0.49\textwidth}
\includegraphics[width=1\linewidth]{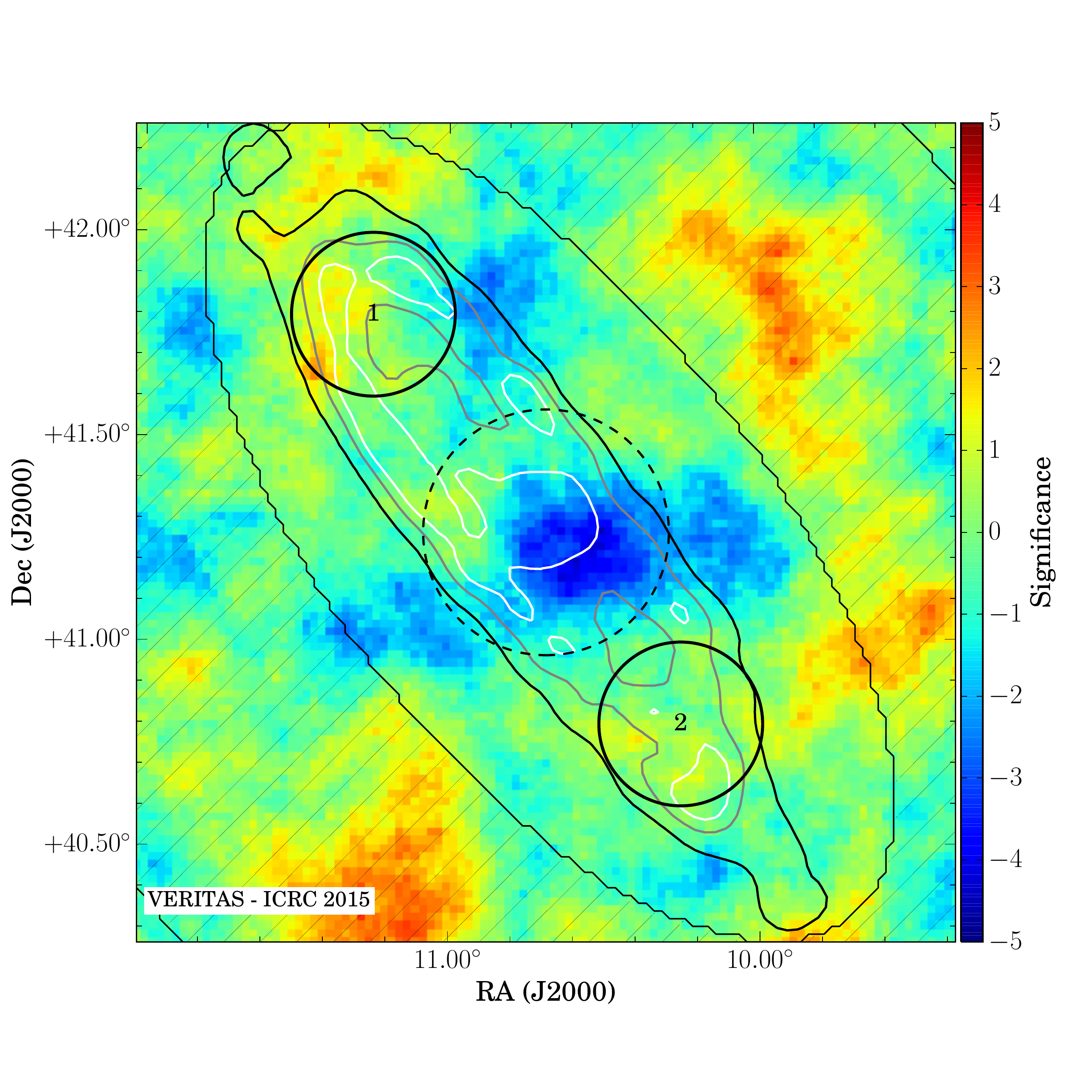}
\caption{test region radius = \ang{0.2} (\textit{Large})}
\label{Fig:VERSkyMapAllHardExtED}
\end{subfigure}
\caption{Skymaps of the complete VERITAS data set on M~31 overlaid with contours from the IRIS \SI{100}{\micro m} map.  The test regions are shown (solid) along with the central exclusion region (dashed) and the background region (hatched).  The impact of the optical brightness on the central region is clearly visible.}
\end{figure}

Using the standard techniques it is not possible to determine the flux from the entire galaxy, instead a method is employed that combines the information from multiple test regions within M~31 and then scales the result to determine the flux from the entire galaxy.
This requires a model of the expected emission, we have used the same template as was found to be the best fit in the \Fermis analysis, the \textit{Raw} model based upon the IRIS \SI{100}{\micro m} map (Figure~\ref{Fig:IRISRaw}).
The galaxy was sampled using two different test region sizes, one of radius \ang{0.1} (\textit{Small}, 8 test positions) and one of radius \ang{0.2} (\textit{Large}, 2 test positions).
The test regions were selected using the same IRIS template to cover as much of M~31 as possible, excluding the bright central region and without any overlap (see Figures \ref{Fig:VERSkyMapAllHardED} and \ref{Fig:VERSkyMapAllHardExtED} for positions).
To determine the excess counts and $\alpha$, the ratio of the acceptances in the signal and background regions, the counts and acceptances from all of the test regions are combined and the background is determined using an elliptical ring surrounding M~31.
The effective areas (which are generated assuming a point source) for each test position (calculated for the pointing, exposure, radial acceptance, etc. for that position) are combined by adding the effective area for each test position after correction for the expected difference in response for the predicted flux (calculated using the \emph{Raw} template folded with the PSF (\ang{0.1}) in comparison with a point source (folded with the same PSF).
This is then converted into a flux for the whole of M~31 by scaling the  flux from the test positions by the fraction of the flux that was expected from those test positions using the \textit{Raw} template.

After a thorough data quality assessment, in particular removing any periods of time affected by clouds or hardware issues, 54.69 hours of data (livetime) was analysed using the standard VERITAS data analysis package.
No evidence for emission from M~31 was detected in any of the previously defined test regions.
 
A 95\% confidence level upper limit is put on the flux from the entire galaxy using bounded Rolke method \cite{Rolke2005} and assuming a spectral index of -2.5 (based upon the \Fermis result and observations of diffuse VHE emission within the MW).
Using the \textit{Small} test locations differential upper limit is \SI{6.91e-15}{GeV^{-1} cm^{-2} s^{-1}} at \SI{416.9}{GeV} (the minimum energy for reconstruction based upon a maximum of 10\% energy bias, significantly higher than the VERITAS energy threshold due to the large \emph{Size} cut employed and the location within the FoV), for the \textit{Large} test locations it is \SI{2.7e-14}{GeV^{-1} cm^{-2} s^{-1}} at \SI{346.7}{GeV}.
Integral upper limits over the range from the minimum safe energy to \SI{30}{TeV} are \SI{1.9e-12}{cm^{-2} s^{-1}} (2.2\% of the Crab Nebula flux) and  \SI{6.2e-12}{cm^{-2} s^{-1}} (5.2\%) respectively.

\section{Discussion \& Conclusions}
M~31 poses many challenges for analysis with IACTs, principally its apparent size and optical brightness, thus requiring special consideration when analysing the data.
54.69 hours of VERITAS observations of M~31 are presented, giving 95\% confidence level upper limits on the total VHE emission from M~31 at \SI{6.9e-15}{GeV^{-1} cm^{-2} s^{-1}} at \SI{416.9}{GeV} (\textit{Small} test regions) and \SI{2.7e-14}{GeV^{-1} cm^{-2} s^{-1}} at \SI{346.7}{GeV} (\textit{Large} test regions) assuming a spectral index of -2.5.
This limit is significantly above any model predictions of the total flux from M~31 (though no detailed modelling of M~31 has been done) and the scaled flux from the starburst galaxies M~82 and NGC~253 but, combined with the lack of any evidence of from  the skymaps, it shows that there are no extremely bright sources within M~31 nor any regions that show anomalous emission.

Six and a half years of \Fermis data shows that emission is likely to be from an extended source.  
At the lowest energies, there is a suggestion of a turnover in the spectrum, indicative of pion emission being the source of the HE \GRs emission.
Future analysis with \textit{Pass 8} will provide significantly greater insight into this.
A simple scaling of GALPROP model of the Milky Way fits the \Fermis data well, at the energies of VERITAS it predicts at least 1.5 orders of magnitude less flux than the upper limit (though the contribution from unresolved point sources is likely to be comparable to that from the diffuse emission at these energies).

\acknowledgments
This research is supported by grants from the U.S. Department of Energy Office of Science, the U.S. National Science Foundation and the Smithsonian Institution, and by NSERC in Canada. 
We acknowledge the excellent work of the technical support staff at the Fred Lawrence Whipple Observatory and at the collaborating institutions in the construction and operation of the instrument.
R. Bird is funded by the DGPP which is funded under the Programme for Research in Third-Level Institutions and co-funded under the European Regional Development Fund (ERDF).
The VERITAS Collaboration is grateful to Trevor Weekes for his seminal contributions and leadership in the field of VHE gamma-ray astrophysics, which made this study possible.

\end{document}